\documentclass[aps,prl,twocolumn,groupedaddress,amsmath,amssymb]{revtex4}

\usepackage{graphicx}
\usepackage{dcolumn}
\usepackage{bm}

\begin{document}
\title{Readout strength dependence of state projection in superconducting qubit}
\author{K. Kakuyanagi$^{1}$, H. Nakano,$^{1}$ S. Kagei,$^{1,2}$ R. Koibuchi,$^{1,2}$ S. Saito,$^{1}$ 
A. Lupa\c{s}cu,$^{3}$ 
K. Semba$^{1}$}
\affiliation{
$^{1}$NTT Basic Research Laboratories, NTT Corporation, Kanagawa, 243-0198, Japan\\
$^{2}$Tokyo University of Science, 1-3 Kagurazaka, Shinjuku, Tokyo 162-8601, Japan\\
$^{3}$Delft University of Technology, Lorentzweg 1, 2628 CJ Delft, The Netherlands
}

\begin{abstract}
We have controlled the measurement strength of the qubit readout process by using a Josephson bifurcation amplifier (JBA : an AC-driven SQUID).
The optimum readout pulse changes the superposition state into an eigenstate.
To clarify how a sufficiently strong measurement causes a projection phenomenon, we have studied the projection conditions in the JBA readout.
We found that the projection occurs above the pulse height at which visibility starts to appear.
Furthermore, the projection continues to occur as readout strength increases, even though the visibility vanishes.
This result helps us understanding the JBA readout process.
\end{abstract}


\maketitle

Projection measurement is very important for quantum information processing applications for example, measurement-based quantum computing \cite{Projection}. 
And it is also important as regards understanding temporal quantum correlation \cite{LGneq}.
Recently, with the help of microfabrication techniques, it has become possible to fabricate controllable artificial quantum systems \cite{SCQ1,SCQ2,SCQ2b,SCQ3,SCQ4}. 
To measure these quantum states, we need to detect small changes in quantum systems.
Techniques for measuring quantum state of solid state qubit have progressed significantly.
For example, a partial measurement \cite{PartialMes} has been reported in the field of superconducting qubits.
They controlled the potential barrier height of a phase qubit and succeeded in observing wavefunction modification using partial measurements.
In addition, successful quantum non-demolition (QND) measurements \cite{QND1,QND2} have been reported.
They carried out multiple quantum measurements during the coherence time and confirmed that the post-measurement state was maintained by using Josephson bifurcation amplifier (JBA) \cite{JBA} measurements.
The JBA technique uses a nonlinear resonator as the qubit quantum state probe.
By choosing the optimum bias condition, the bifurcation phenomenon in a strongly driven nonlinear resonator, which has Josephson junctions, becomes sensitive to the coupled quantum system.
Therefore we can detect the quantum state of a qubit.

When the detector has a lot of degrees of freedom, it is very difficult and sometimes impossible to trace the dynamics during the measurement.
A JBA measurement is an ideal quantum probe because it realizes a QND measurement.
The JBA readout forms the relation between the qubit state and the final stable state (low- or high-amplitude oscillation state) of the JBA resonator.
Very fortunately our detector JBA is a simple system, and the interaction between the JBA resonator and a superconducting qubit is well known.
Recently, we have made some progress on a theoretical analysis of the JBA readout \cite{JBAana}. 
We are interested in determining the condition that induces the projection in a real JBA measurement.
In our system, it is possible to program the strength of the quantum detection.
So we changed the measurement strength and clarified the projection condition.

To readout the qubit state, we employ a short JBA pulse (hereafter called a readout pulse).
It excites the oscillation in the JBA resulting in the generation of an interaction between the JBA and the qubit.
In the following, we describe our experimental findings as regards the quantum detection process with a JBA.

\begin{figure}[tbp]
\begin{center}
\includegraphics[width=1.0\columnwidth]{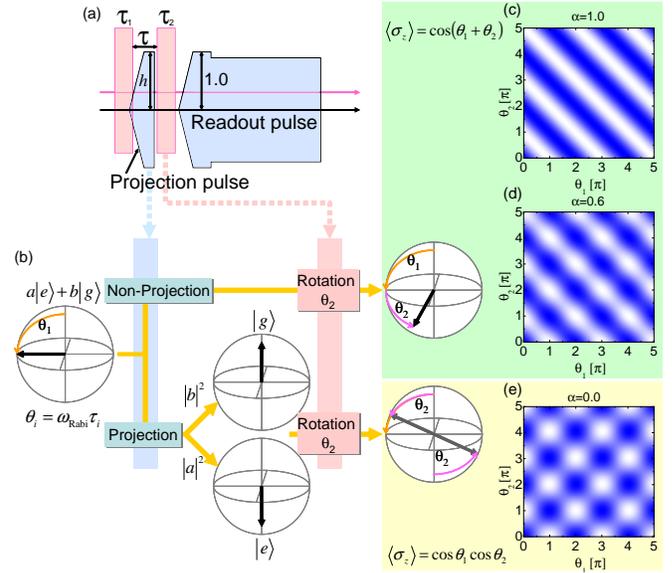}
\caption{
 (a) Pulse sequence of the projection measurement. 
The second control pulse works as a rotation gate.
Between the two control pulses, we apply a projection pulse from the JBA transmission line.
After the second control pulse, we apply a readout pulse to read the qubit state.
(b)
Schematic images of the qubit state along the measurement sequence.
The left Bloch sphere shows the quantum superposition state just after the first control pulse.
The upper (lower) right Bloch spheres are the qubit states just before the readout pulse in the case of non-projection (projection).
After a projection, the Bloch vector is rotated by the second control pulse, so $\langle \sigma_z \rangle$ will be $\langle \sigma_z \rangle = \cos{\theta_1}\cos{\theta_2}$.
(c,d,e) shows the calculated $\langle \sigma_z \rangle$ patterns expected with the  pulse sequence for three $\alpha$ values
}
\label{FIG_A}
\end{center}
\vspace{-0.2cm}
\end{figure}
The front part of the readout pulse plays an essential role in the qubit readout.
So, to examine how the JBA readout process occurs, we observed the changes in the resulting qubit state when we applied various projection pulses with the same shape as the front part of our readout pulse.
To examine the projection conditions of the superposition state, we employed a pulse sequence that consisted of two successive qubit control pulses with a short projection pulse between them (Figure \ref{FIG_A}(a)).
The qubit is initialized in the ground state simply by allowing it to relax.
By applying the first control pulse we generated a superposition state of our qubit ($|\Psi\rangle = \sin{\theta_1 \over 2} |e \rangle + \cos{\theta_1 \over 2}|g\rangle$).
Then, we applied the projection pulse to the JBA.
After that, to observe the state change that it caused, we applied a rotation operation to this qubit state using the second control pulse.
If the projection pulse does not induce a projection, the first control pulse should induce qubit Bloch vector rotation by $\theta_1$, and in addition the second control pulse should rotate it by $\theta_2$.
If we can neglect the phase relaxation in the qubit, the expected value of $\sigma_z$ becomes $\langle \sigma_z \rangle = \cos{\left(\theta_1+\theta_2\right)}$ (Figure \ref{FIG_A}(b) top), where $\sigma_z=|g\rangle\langle g|-|e\rangle\langle e|$ is the Pauli operator corresponding to the qubit energy eigenstate.
On the other hand, if a projection occurs through the application of the projection pulse, the $\langle \sigma_z \rangle$ immediately after the projection becomes $\cos{\theta_1}$. 
After applying the second control pulse to this state, $\langle \sigma_z \rangle$ reads $\cos{\theta_1}\cos{\theta_2}$ (Figure \ref{FIG_A}(b) bottom).
We can represent $\langle \sigma_z \rangle$ in more detail as follows.
\begin{equation}
\langle \sigma_z \rangle = \cos{\theta_1} \cos{\theta_2} -  \alpha \sin{\theta_1}\sin{\theta_2}
\label{Sigma}
\end{equation}
In this formula $\alpha$ is an indicator of projection.
By calculating the time evolution of $\left<\sigma_z\right>$ in detail, we can describe $\alpha$ in equation (\ref{Sigma}) as follows.
\begin{equation}
\alpha = \alpha_0 e^{-{\tau\over T_2}} \cos{\left(\left(\omega-\omega_0\right)\tau \right)}
\label{alpha}
\end{equation}
So, $\alpha$ is described as a product of a bare projection indicator $\alpha_0$, a dephasing part and a detuning part.
In the absence of dephasing and detuning (that is, $T_2 \rightarrow \infty$, $\omega \rightarrow \omega_0$), $\alpha$ is identical to $\alpha_0$.
When no projection occurs, $\alpha$ takes a finite value.
On the other hand, when projection occurs, $\alpha$ becomes 0 regardless of dephasing or detuning.
By applying a projection pulse within the qubit coherence time, we can detect the ``wavepacket reduction" caused by the projection pulse.
Figure \ref{FIG_A}(c,d,e) shows the calculated $\langle \sigma_z \rangle$.
In the absence of any projection and any dephasing, the $\alpha$ value of equation (\ref{Sigma}) becomes 1.
Therefore, we obtain a stripe pattern in the $\langle \sigma_z \left(\theta_1, \theta_2\right)\rangle = \cos{\left(\theta_1+\theta_2\right)}$ plot (Figure \ref{FIG_A}(c)).
When a projection is caused by the applied projection pulse, $\alpha$ becomes 0, so the checkerboard pattern seen in Figure \ref{FIG_A}(e) is expected.

\begin{figure}[tbp]
\begin{center}
\includegraphics[width=1.0\columnwidth]{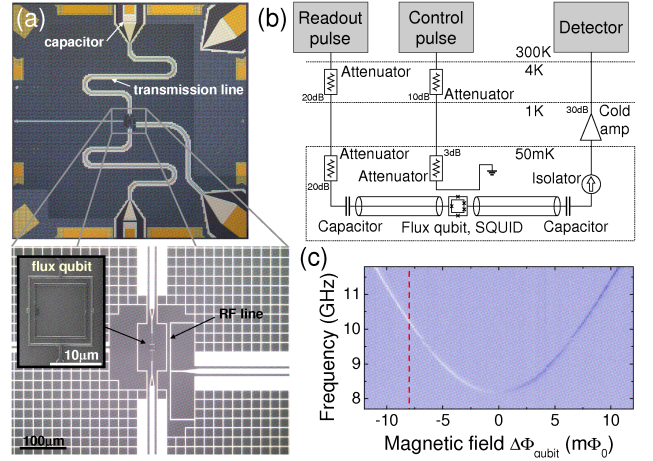}
\caption{
(a) Image of the sample for JBA readout. We designed a SQUID structure at the center of a $\lambda/2$ coplanar transmission line resonator. The flux qubit is magnetically coupled with the SQUID. The mutual inductance of this coupling is $\sim$ 14 pH. To control the qubit state we installed on RF line near the qubit. The mutual inductance between the qubit and the RF line is $\sim$ 0.1 pH.
(b) Schematic diagram of the measurement setup.
(c) shows the qubit spectrum of the sample.
The qubit gap frequency is $\Delta = 8.1$ GHz and the persistent supercurrent away from the degeneracy point is $Ip = 130$ nA.
The broken red line shows the magnetic field in which we performed the projection measurement.
}
\label{FIG_B1}
\end{center}
\vspace{-0.2cm}
\end{figure}

In our JBA, the characteristic bifurcation time (the transition from the initial state to the final state) depends on the $Q$ factor and the resonance frequency $\omega_{\rm JBA}$, as ${Q \over \omega_{\rm JBA}}$.
The resonance frequency of the JBA is approximately 7.5 GHz.
The $Q$ factor of our JBA resonator is approximately 300, so the characteristic bifurcation time is $\sim 6$ ns.
We kept our flux qubit under an external magnetic field of $\Delta\Phi_{\rm qubit} = \Phi_{\rm qubit}- {2n+1\over 2}\Phi_0$= -8 ${\rm m}\Phi_0$, where $\Phi_0={h\over 2e}$ : the flux quantum.
With this magnetic field, the resonance frequency of the qubit is 10.3 GHz.
We used a dilution refrigerator, and measured the sample at temperatures below $50$ mK.
Figure \ref{FIG_B1}(b) is a schematic of our measurement setup.
A JBA readout pulse and the projection pulse propagate along the transmission line resonator in the refrigerator, and are amplified by a low noise cold amplifier.
Which resonance state of the JBA is realized by the readout pulse depends on the qubit state, so we can readout the qubit state by measuring the amplitude and phase of the obtained output signal of the amplifier using a homodyne detection technique.
By applying a resonant microwave pulse to the control line, we can control the qubit state before the projection pulse.

We employed the sequence shown in Figure \ref{FIG_A}(a) and observed the way in which the projection of a superposition state takes place.
When we read out the qubit state, the JBA resonator bifurcates into a high or low amplitude state, and these two states correspond to the qubit ground state and the qubit excited state, respectively.
To obtain $\langle \sigma_z \rangle$, we observed the probability of the JBA low amplitude state $P$.
The relationship between $\langle \sigma_z \rangle$ and $P$ can be described as follows:
$\langle \sigma_z \rangle = 2 \left\{\left(P_0-P\right)/V+P_0\right\}-1$.
In this formula $P_0$ is an offset and $V$ is visibility in the measured probability.
We chose the distance between the two control pulses $\tau=12$ ns to reduce the influence of the phase relaxation (Figure \ref{FIG_A}(a)).
At this flux bias, the energy relaxation time of the qubit is $T_1 = 58$ ns, and the phase relaxation time is $T_2 = 13$ ns. 
Because both the projection process and the phase relaxation contribute to the reduction of the $\sin{\theta_1}\sin{\theta_2}$ term in equation (\ref{Sigma}), a short $\tau$ is better for this projection measurement.
And we optimized the amplitude and frequency of the applied JBA pulses to read out the qubit state.
The rise time of the projection pulse was about 10 ns, and it maintained a constant height for 10 ns.
We varied the pulse lengths of both control pulses from 0.5 to 10 ns.
After the second control pulse, we detected the qubit state using a JBA readout pulse.
By averaging typically 20000 times, we obtained probability $P$.

\begin{figure}[tbp]
\begin{center}
\includegraphics[width=0.9\columnwidth]{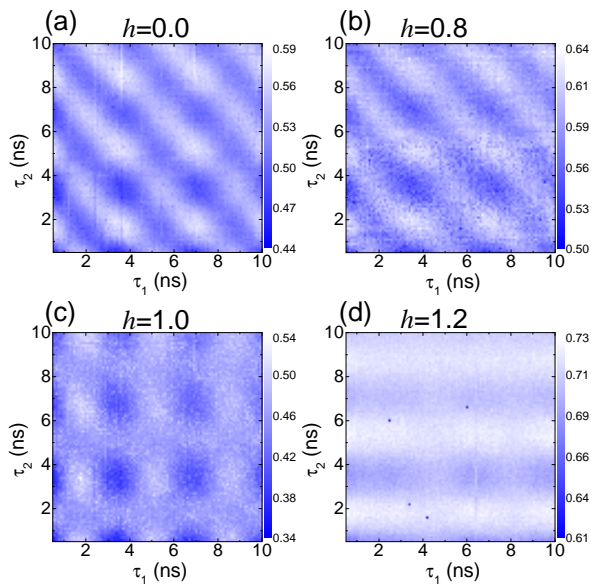}
\caption{
The observed probability $P\left(\tau_1,\tau_2\right)$ patterns, which are related to $\langle \sigma_z \rangle$, with various projection pulse heights $h$. 
In this experiment, the $\pi$ pulse length was 1.75 ns ($\theta_i=\omega_{\rm Rabi} \tau_i$, here $\omega_{\rm Rabi}$ is the Rabi frequency).
}
\label{FIG_B2}
\end{center}
\vspace{-0.2cm}
\end{figure}
Figure \ref{FIG_B2} shows the observed $P\left(\tau_1,\tau_2\right)$ pattern we obtained when we employed projection pulses with different heights.
$h$ is the pulse height normalized by that of the readout pulse usually used in our JBA measurements.
When we employed an $h=0.8$ pulse, we observed the same pattern as without a projection pulse ($h=0.0$) namely a stripe pattern, so the projection did not occur.
However, when we used an $h=1.0$ pulse, the $\sin{\theta_1} \sin{\theta_2}$ component of equation (\ref{Sigma}) vanished and a clear checkerboard pattern appeared (Figure \ref{FIG_B2}(c)).
This sudden change from the stripe pattern to the checkerboard pattern suggests that the projection was induced by the applied pulse although the pattern contrast was not as good as the calculated result because of the energy relaxation.

When we further increase the pulse height ($h=1.2$), the observed pattern does not depend on the first control pulse, because the quantum state is destroyed by the over-strong projection pulse (Figure \ref{FIG_B2}(d)).
However we observed a small Rabi oscillation caused by the second control pulse.
This is due to the energy relaxation that occurs after the randomization of the qubit state, which induces a bias in the post-pulse state.

To discuss the observed projection pulse height dependence quantitatively, we use $\alpha$.
$\langle \sigma_z \left(\theta_1,\theta_2\right)\rangle$ obtained from the sequences shown in Figure \ref{FIG_A}(a) corresponds to a function of the rotation angle caused by the first control pulse $\theta_1=\omega_{\rm Rabi} \tau_1$ and the second control pulse $\theta_2 = \omega_{\rm Rabi} \tau_2$ where $\omega_{\rm Rabi}$ is the Rabi frequency for the control pulses.
So, from equation (\ref{Sigma}), $\alpha$ was estimated as follows:
$\alpha = {1\over 4} \left\{ \langle \sigma_z \left({3\pi\over 2}, {\pi \over 2} \right)\rangle-\langle \sigma_z \left({\pi\over 2}, {\pi \over 2} \right)\rangle+\langle \sigma_z \left({\pi\over 2}, {3\pi \over 2} \right)\rangle-\langle \sigma_z \left({3\pi\over 2}, {3\pi \over 2} \right)\rangle \right\}$.

\begin{figure}[tbp]
\begin{center}
\includegraphics[width=0.9\columnwidth]{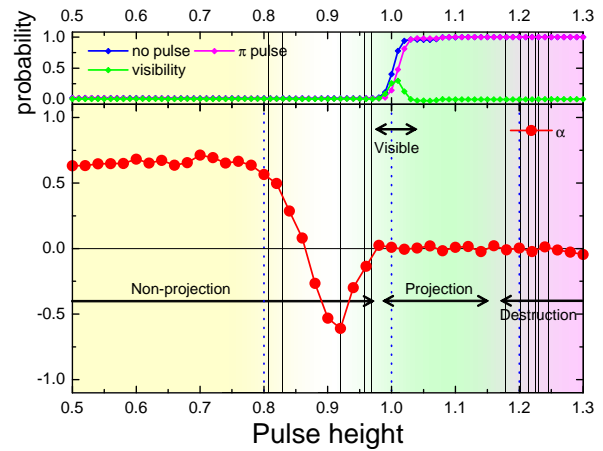}
\caption{
Pulse height dependence of the S-curve (probability with which we find a low state JBA) and pulse height dependence of $\alpha$.
The magenta and blue lines are S-curves for the qubit excited state and ground state, respectively.
The green line is the difference between these two S-curves, which corresponds to the visibility of the qubit.
The filled red circles show the projection pulse height dependence of the $\alpha$ indicator.
}
\label{FIG_C}
\end{center}
\vspace{-0.2cm}
\end{figure}
Figure \ref{FIG_C} shows the dependence of $\alpha$ on the projection pulse height $h$.
The top panel shows S-curves, which reveal that our readout has a finite visibility in the $0.98 < h < 1.03$ region.
This means that our readout can provide qubit state information only in this region.
Below $h=0.8$, $\alpha$ has a positive value of 0.65.
Because of the small dephasing that occurs during in the pulse sequences, $\alpha$ became less than 1.
No projection occurs in the region.
We tuned the frequency of an applied control pulse to the energy of the flux qubit ($\omega=\omega_0$).
Therefore, $\alpha$ should have a positive value.
However, above $h=0.8$, $\alpha$ decreases and has negative values.
In this region, projection does not occur but the excited photon number in the JBA resonator increases rapidly.
The interaction between the qubit and the JBA depends on the photon number in the JBA resonator, so the effective energy of the flux qubit deviates from $\omega_0$ to $\omega^{\prime}_0$ when the projection pulse is applied.
Because of this energy shift, the phase of the qubit state rotates during the period of the projection pulse, and $\alpha$ adopts negative values as $\cos{\left(\left(\omega-\omega^{\prime}_0\left(h\right)\right)\tau\right)}< 0$ (see eq. \ref{alpha}).
It should be noted that a finite (negative) $\alpha$ means the absence of projection.
The absolute amplitude of $\alpha$ at $h=0.92$ is almost the same as that of $\alpha$ below $h=0.8$.
This means that the application of the projection pulse itself does not induce significant decoherence (dephasing) at least below $h=0.92$.
Above $h=0.98$, $\alpha$ becomes 0 and this suggests that the projection of the qubit state has appeared.
In the $0.98<h<1.03$ region, the resonance state of the JBA bifurcates and different qubit states lead the JBA to different final states.
So we can detect the qubit state by observing the resonance state of the JBA resonator.
Above $h=1.03$, we still observe the checkerboard pattern of $\langle \sigma_z\rangle$, which is the same as in Figure \ref{FIG_B2}(c). This shows that the projection pulse has induced projection. However, because of the large pulse amplitude, the JBA reaches the same final resonance state, independent of the state into which the qubit is projected.
Therefore, in this region, qubit state projection occurs although the visibility of the JBA measurement is 0.
In a simple description of the measurement, the projection event and the possibility of obtaining qubit state information has a one to one relationship.
However, here ($1.03<h<1.2$) the post-measurement state of the qubit is a projected state although we cannot obtain any information about the qubit from the JBA readout.
This is a clear example of a fact that such a simple relationship is not valid in a real measurement where we have to take account of the dynamics of the detector.
Above $h=1.2$, the $\theta_1$ dependence of $\langle \sigma_z \rangle$ vanishes (Figure \ref{FIG_B2}(d)). This behavior means that the qubit state is destroyed by the over-strong projection pulse.

The above results show that the normal JBA readout ($h=1.0$) method projects the qubit state to one of the energy eigenstates and maintains that state.
The JBA readout is almost an ideal quantum state measurement method, and it also realizes a quantum non-demolition measurement, where the post-measurement state is kept in the projected state.

We succeeded in observing the behaviors in projection measurements of superconducting qubit states by using a transmission line type JBA as a detector.
When we start to drive the JBA with a readout pulse (its front is the projection pulse, examined above), the qubit and JBA begin to interact and form correlations (entanglement). However, the application of a pulse smaller than the threshold ($h\sim 0.98$) does not cause a projection or dephasing in the qubit. When the pulse height becomes large enough to make the JBA states for two possible qubit states distinguishable, projection occurs and the possible qubit-JBA composite system becomes one of two possible classically correlated qubit-JBA states. Then, the quantum measurement has been accomplished. When the pulse height is in the range where only one of the qubit-JBA states performs the JBA transition, we can detect the qubit state from the JBA readout.
Our results clarified experimentally the way in which the qubit state projection to an eigenstate is caused by the JBA bifurcation phenomenon.
This supports a previous theoretical analysis of the JBA readout \cite{JBAana}, and is an important result for understanding the mechanisms of quantum state measurement.

We thank Professor Hideaki Takayanagi of Tokyo University
of Science and NIMS for his essential support throughout this work. 
This work was supported in part by Grant-in-Aid for Scientific Research of Specially Promoted Research \#18001002 by MEXT, and Grant-in-Aid for Scientific Research (A) \#18201018 by JSPS.

\end{document}